\documentclass[pre]{revtex4}
\usepackage[english]{babel}
\usepackage{amsmath}
\usepackage{epsfig}

\begin{document}

\title{\bf IMPOSSIBILITY OF THE EXISTENCE OF THE UNIVERSAL DENSITY FUNCTIONAL
}
\author{V.B. Bobrov $^{1}$, S.A. Trigger $^{1,2}$}
\address{$^1$ Joint\, Institute\, for\, High\, Temperatures, Russian\, Academy\,
of\, Sciences, Izhorskaia St., 13, Bd. 2. Moscow\, 125412, Russia;\\
emails:\, vic5907@mail.ru,\;satron@mail.ru\\
$^2$ Eindhoven  University of Technology, P.O. Box 513, MB 5600
Eindhoven, The Netherlands}

\begin{abstract}
Using the virial theorem, it is shown that the hypothesis of the
existence of the universal density functional is invalid. \\

PACS number (s): 31.15.E-, 71.15.Mb, 52.25.-b, 05.30.Fk

\end{abstract}

\maketitle

The density functional theory (DFT) is based on two statements
formulated in the known paper by Hohenberg and Kohn [1]. According
to the first statement conventionally referred to as the
Hohenberg-Kohn theorem, the ground state energy of the
inhomogeneous electron gas in a static external field is the
functional o the inhomogeneous electron density. The
Hohenberg-Kohn theorem proof is based on the lemma [1,2]: the
inhomogeneous density $n({\bf r})$  in the ground state of a bound
system of interacting electrons in a certain static external field
characterized by the potential $\varphi^{ext}({\bf r})$ uniquely
defines this potential. In this case, according to [1,2]: (i) here
the term "uniquely" means "with an accuracy" to an additive
constant of no interest; (ii) in the case of a degenerate ground
state, the lemma relates to the density  $n({\bf r})$ of any
ground state. The requirement of the ground state non-degeneracy
is easily eliminated [3]; (iii) the lemma is mathematically
rigorous.

According to the lemma, the inhomogeneous density $n({\bf r})$
corresponding to the ground state and the external field potential
$\varphi^{ext}({\bf r})$ are biunique functionals,

\begin{eqnarray}
\varphi^{ext}({\bf r})=\varphi^{ext}(\{n {(\bf
r})\})\leftrightarrow n({\bf r})=n(\{\varphi({\bf r})\}),
\label{G1}
\end{eqnarray}
Thus, for the ground state energy $E_0$  of the system of
interacting electrons with Hamiltonian $H$  in an external field
with potential $\varphi^{ext}({\bf r})$, characterized by the wave
function $\mid\Psi_0\rangle$ , we can write
\begin{eqnarray}
E_0=\langle \Psi_0 |H| \Psi_0 \rangle=E_0 (N,\{\varphi({\bf
r})\})=E_0 (\{n({\bf r})\}), \label{G2}
\end{eqnarray}
Let us pay attention that the lemma is a direct consequence of the
Rayleigh-Ritz minimum principle which, as applied to quantum
mechanics, is written as (see, e.g., [4,5])
\begin{eqnarray}
E_0\leq \langle \Psi |H| \Psi \rangle, \qquad\, \langle \Psi\mid
\Psi \rangle=1 \label{G3}
\end{eqnarray}
where  $\mid \Psi\rangle$ is any normalized wave function for the
system of $N$ electrons. In non-strict inequality (3), equality
takes place only in the case when $\mid \Psi\rangle=\mid
\Psi_0\rangle$. However, direct application of variational
inequality (3) to the DFT is impossible, since this inequality
implies variation of the wave function $\mid\Psi\rangle$, rather
than the density $n({\bf r})$. The point is that, according to the
Hohenberg-Kohn theorem, only the ground state energy is a density
functional. To obtain a variational inequality similar to (3), but
implying the possibility of varying the inhomogeneous density,
Hohenberg and Kohn [1] formulated the second statement that the
quantity  $\langle \Psi_0 |T+U| \Psi_0 \rangle$ is a universal
functional of density $n({\bf r})$,
\begin{eqnarray}
F^{univ} (\{n({\bf r})\})=\langle \Psi_0 |T+U| \Psi_0 \rangle=E_0
(\{n({\bf r})\})-\int \varphi^{ext}({\bf r}) n({\bf r})d {\bf r}
\label{G4}
\end{eqnarray}
where $T$  and $U$  are the operators of the kinetic energy and
interparticle interaction energy of electrons, respectively. In
this case, the universality is understood as the functional
independence of the form and value of the external field
$\varphi^{ext}({\bf r})$ . In contrast to the Hohenberg-Kohn
theorem, the second statement (4) is accepted in the DFT without
proof. Moreover, the exact form of this universal functional is
still unknown even for noninteracting particles ($U=0$). However,
the use of the assumption on the existence of such a universal
functional makes it possible to "create" a variational procedure
for determining the inhomogeneous density $n({\bf r})$
corresponding to the ground state of the system under
consideration. The essence of this procedure is reduced to the
following: let the explicit form of the density functional for
ground state energy $E_0 (\{n({\bf r})\})$  be a priori known,
whereas the inhomogeneous density $n({\bf r})$  corresponding to
the ground state is unknown. Then, to determine $n({\bf r})$,
according to (4), the inequality [1,2]
\begin{eqnarray}
E_0 (\{n({\bf r})\})\leq E_0 (\{\tilde n({\bf r})\}) \label{G5}
\end{eqnarray}
should be used,  where $\tilde n({\bf r})$   is any positive
function satisfying the condition $\int \tilde n({\bf r}) d {\bf
r}=1$. In this case, equality in (5) takes place if $\tilde n({\bf
r})=n({\bf r})$ . Thus, the DFT loses practical meaning without
statement (4), since the use of only the Hohenberg-Kohn theorem
does not "relive the need" to find a many-body wave function to
determine the inhomogeneous density.

Let us now show that statement (4) is invalid. We perform the
proof "to the contrary" by analogy with the proof of the lemma
[1,2]. Indeed, if the functional $F^{univ} (\{n({\bf r})\})$ is a
universal functional of the inhomogeneous density $n({\bf r})$,
its variational derivative with respect to the density is also a
universal functional,
\begin{eqnarray}
G^{univ} (\{n({\bf r})\})=\frac{\delta F^{univ} (\{n({\bf
r})\})}{\delta n({\bf r})} \label{G6}
\end{eqnarray}
Substituting the last equality on the right-hand side (4) into
(6), and taking into account that $\langle\delta \Psi_0 \mid H\mid
\Psi_0\rangle+\\
\langle \Psi_0 \mid H\mid \delta \Psi_0\rangle=0$
we find
\begin{eqnarray}
G^{univ} (\{n({\bf r})\})=\frac{\delta E_0 (\{n({\bf
r})\})}{\delta n({\bf r})}-\varphi^{ext}(\{n {(\bf r})\})-\int
n({\bf r}_1)\frac{\delta \varphi^{ext}(\{n {(\bf r}_1)\})}{\delta
n({\bf r})}d {\bf r}_1 \label{G7}
\end{eqnarray}
Taking into account (1) and (2) one arrive at
\begin{eqnarray}
\frac{\delta E_0 (\{n({\bf r})\})}{\delta n({\bf r})}=\int
\frac{\delta E_0 (N,\{ \varphi^{ext}({\bf r})\})}{\delta
\varphi^{ext}({\bf r}_1)}\frac{\delta \varphi^{ext}({\bf
r}_1)}{\delta n({\bf r})}d {\bf r}_1, \qquad  \frac{\delta E_0
(N,\{ \varphi^{ext}({\bf r})\})}{\delta \varphi^{ext}({\bf
r})}=n({\bf r}) \label{G8}
\end{eqnarray}
It follows from (7) and (8) that
\begin{eqnarray}
G^{univ} (\{n({\bf r})\})= -\varphi^{ext}(\{n({\bf
r})\})\label{G9}
\end{eqnarray}
Thus, if the quantity $F^{univ} (\{n({\bf r})\})$  is a universal
density functional, the external field potential is also a
universal functional of the inhomogeneous density. Furthermore,
according to (4), the ground state energy $E_0$ is also a
universal functional of the inhomogeneous density. These
statements are easily refuted using the virial theorem. Indeed, in
the case of finite motion of the system of noninteracting ($U=0$)
particles in the external field $\varphi^{ext}({\bf r})$,
according to the virial theorem (see [6] and references therein),
we have
\begin{eqnarray}
2\langle \Psi_0 |T| \Psi_0 \rangle - \int n({\bf r}) \left({\bf
r}\cdot \nabla\varphi^{ext}({\bf r})\right) d {\bf r}=0\label{G10}
\end{eqnarray}
If a homogeneous coordinate power function $\varphi^{ext}({\bf
r})\sim r^m$   can be chosen as a particular case for the external
field potential, then
\begin{eqnarray}
\langle \Psi_0 |T| \Psi_0 \rangle=\frac{m}{2} \int
\varphi^{ext}({\bf r}) n({\bf r}) d {\bf r}, \qquad\,
E_0=\left(\frac{m}{2}+1\right) \int \varphi^{ext}({\bf r}) n({\bf
r}) d {\bf r} \label{G11}
\end{eqnarray}
Hence, if the external field potential is a universal functional
(9), both quantities in (11) are not universal density functionals
(they depend on the field parameter $m$), and we come to
contradiction. Thus, the hypothesis of the existence of the
universal density functional (4) is invalid. We note that the
doubts on the possible existence of the universal density
functional were probably first casted by Gilbert [7], but in this
paper it was only an assumption, related solely with consideration
of the nonlocal type of an external field.

Therefore, the development of the density matrix functional theory
(DMFT) becomes particularly urgent (see, e.g., [8,9] and
references therein). At the same time, the DMFT is consistent with
the virial theorem, including the case of the consideration of the
volume occupied by the system [10]. We note that on the basis of
the virial theorem appears the way for introduction of the really
universal DMFT.

\section*{Acknowledgment}

This study was supported by the Netherlands Organization for
Scientific Research (NWO), project no. 047.017.2006.007 and the
Russian Foundation for Basic Research (the project no.
10-02-90418-Ukr-a).

\end{document}